\documentstyle[aas2pp4]{article}
\input{epsf}

\def\kms{km s$^{-1}$} 
\def\etal{{\it et al.}}

\def\imf{{\sc imf\/}}
\def\re{{r_e}}
\def\mue{{\mu_e}}
\def\zform{z_{form}}

\def\Sec{${}^{\prime\prime}$\llap{.}}

\def\rq{$r^{1/4}$-law}

\def\etal{{\it et~al.\/}}

\def\kms{{km~s$^{-1}$}}
\def\kpc-1{{kpc$^{-1}$}}
\def\Mpc-1{{Mpc$^{-1}$}}
\def\s-1{{sec$^{-1}$}}
\def\pdeg2{{deg$^{-2}$}}

\def\h0{{H$_0$}}
\def\q0{{$q_0$}}

\lefthead{Kelson \etal}

\begin{document}

\title{Evolution of Early-Type Galaxies in Distant Clusters: The
Fundamental Plane using HST Imaging and Keck Spectroscopy$^{1,2}$}

\author{Daniel D.  Kelson}

\affil{University of California Observatories / Lick Observatory, 
Board of Studies in Astronomy and Astrophysics, 
University of California, Santa Cruz, CA 95064}

\author{Pieter G. van Dokkum, Marijn Franx}

\affil{Kapteyn Astronomical Institute, P.O. Box 800, NL-9700 AV, Groningen,
The Netherlands}

\author{Garth D. Illingworth}

\affil{University of California Observatories / Lick Observatory, 
Board of Studies in Astronomy and Astrophysics, 
University of California, Santa Cruz, CA 95064}

\and

\author{Daniel Fabricant}
\affil{Harvard-Smithsonian Center for Astrophysics, 60 Garden Street, Cambridge,
MA \  02318}

\altaffiltext{1}{Based on observations obtained at the W. M. Keck 
Observatory, which is operated jointly by the California Institute 
of Technology and the University of California.}

\altaffiltext{2}{Based on observations with the NASA/ESA {\it Hubble
Space Telescope}, obtained at the Space Telescope Science Institute,
which is operated by AURA, Inc., under NASA contract NAS 5--26555.}

\begin{abstract}

We present new results on the Fundamental Plane of galaxies in two rich
clusters, Cl1358+62 at $z=0.33$ and MS2053--04 at $z=0.58$, based on Keck
and HST observations.  Our new data triple the sample of galaxies with
measured Fundamental Plane parameters at intermediate redshift.  The
early-type galaxies in these clusters define very clear Fundamental Plane
relations, confirming an earlier result for Cl0024+16 at $z=0.39$. This
large sample allows us to estimate the scatter reliably. We find it to be
low, at 0.067 dex in $\log r_e$, or 17\% in re, similar to that observed in
comparable low redshift clusters. This suggests that the structure of the
older galaxies has changed little since $z=0.58$.

The $M/L_V$ ratios of early-type galaxies clearly evolve with redshift; the
evolution is consistent with $\Delta \log M/L_V \sim -0.3z$.  The $M/L_V$
ratios of two E+A galaxies in Cl1358+62 are also lower by a factor of $\sim
3$, consistent with the hypothesis that they underwent a starburst 1 Gyr
previously.  We conclude that the Fundamental Plane can therefore be used
as a sensitive diagnostic of the evolutionary history of galaxies. Our
data, when compared to the predictions of simple stellar population models,
imply that the oldest cluster galaxies formed at high redshift ($z > 2$).
We infer a different evolutionary history for the E+A galaxies, in which a
large fraction of stars formed at $z<1$. Larger samples spanning a larger
redshift range are needed to determine the influence of starbursts on the
general cluster population.

\end{abstract}

\keywords{galaxies: clusters: general,
galaxies: clusters: individual (Cl1358+62, MS2053--04),
galaxies: evolution,
galaxies: fundamental parameters,
galaxies: kinematics and dynamics}


\section {Introduction}

Evidence for evolution in galaxies at intermediate redshifts has been
found in a number of pioneering studies, both in clusters (e.g.,
Butcher \& Oemler 1978, 1984; Dressler \& Gunn 1983, Couch \&
Sharples 1987) and in the field (e.g., Kron 1980, Broadhurst, Ellis \&
Shanks 1988).  Large spectroscopic surveys with 4 m class telescopes,
coupled with HST images, have begun to clarify the nature of this
evolution. The CFRS redshift survey of $\sim 10^3$ field galaxies (Le
F\'evre \etal\ 1995; Lilly \etal\ 1995) is one such example. Here we pursue
a complementary approach involving the detailed structural and kinematic
study of smaller samples of individual galaxies using HST images and higher
resolution spectroscopy with the Keck telescope.  With these data, we can
exploit the power of the Tully-Fisher relation for spirals ({\it e.g.\/},
Vogt \etal\ 1996) and the Fundamental Plane relation for early-type (E/S0)
galaxies (see Franx 1993).

The Fundamental Plane (Djorgovski \& Davis 1987; Dressler \etal\ 1987)
is particularly valuable due to its low intrinsic scatter (J\o rgensen,
Franx and Kj\ae rgaard 1993). For the Coma cluster, the Fundamental Plane
relation,
\begin{equation}
r_e \propto \sigma^{1.24} I_e^{-0.82},
\label{eq:fp}
\end{equation}
where $r_e$ is the effective radius, $I_e$ is the surface brightness
at that effective radius in the visible, and $\sigma$ is the central
velocity dispersion, has a scatter of only 17\% {\it rms\/} (J\o
rgensen, Franx and Kj\ae rgaard 1996 [JFK96]). This low scatter
implies that the following well-defined relation exists for early-type
galaxies (Faber \etal\ 1987):
\begin{equation}
M/L_V \propto r_e^{0.22}\sigma^{0.49}\propto M^{0.24} r_e^{-0.02}.
\label{eq:ml}
\end{equation}
Thus, the Fundamental Plane is valuable because it explicitly incorporates
galaxy masses. Franx (1995) and van Dokkum and Franx (1996 [vDF]) have
demonstrated the value of this relation for studying evolution in
early-type galaxies at intermediate redshift. The latter authors showed
that the Fundamental Plane in the rich cluster Cl0024+16 is well defined,
and consistent with simple evolutionary models, but the observed sample was
very small.

Here we present new results for two additional clusters at intermediate
redshifts. These new data triple the galaxy sample, and extend the observed
redshift range to $z=0.58$.


\section{Observations and Data Reduction}

The spectroscopic sample was selected on the basis of $R$-band magnitude.
Blue galaxies were rejected to avoid field contamination, though the
color restriction was chosen such that star-forming and post-starburst
cluster members were not excluded. 

Slit masks were designed to include as many bright galaxies as possible,
though we only present data here for galaxies with HST imaging (see Figure
\ref{mosaic} [Plate 1]). In addition, two known ``E+A'' (Dressler \& Gunn
1983) galaxies were added to the Cl1358 mask. These galaxies are not
included in the general sample, but are discussed separately. Thus, ten
galaxies in Cl1358, and five in MS2053, are analyzed in this paper.

\subsection{Spectroscopy}

The spectroscopic observations were made using multi-slit masks with the
Low Resolution Imaging Spectrograph (LRIS) at the Keck telescope.  We
observed at a typical resolution of $\sigma_{instr} = $ 60-85 \kms. The
data reduction was very similar to the data reduction of vDF for Cl0024.
The resulting spectra were very high S/N (typically 20-60 per resolution
element).  We show spectra of two galaxies in Figure \ref{mosaic}(c,d).

We modeled the spectral resolution of the spectrograph in great detail, for
the template stars as well as the galaxies. This is necessary to ensure
that the template stars used for the determination of the velocity
dispersions have the correct spectral resolution. This procedure is the
most essential technical aspect of measuring velocity dispersions of
galaxies at intermediate redshift.

Some galaxies showed peculiar features in their spectra. These features,
either strong Balmer absorption lines, emission lines, residual sky
lines, or atmospheric absorption bands, were given zero weight in the
template fitting. We corrected the central velocity dispersions to an
aperture of 3\Sec 4 at the distance of Coma, using the procedure of
J\o{}rgensen, Franx, \& Kj\ae{}rgaard (1995b). The corrections are small,
1.065 for Cl1358+62 and 1.066 for MS2053--04 ($q_0=0.05$). The resulting
velocity dispersions and random errors are listed in Table 1.

\subsection{Imaging}

We used WFPC2 HST images to measure the structural parameters. Observations
were taken in the filters F606W and F814W for Cl1358+62, and in F702W and
F814W for MS2053--04. These data were processed in the usual way for cosmic
rays and removal of the sky background. The field for Cl1358+62 was very
large, $\sim 8' \times8'$ (Franx \etal\ 1997). For MS2053-04, only 1
central pointing was available. As a result we have more Fundamental Plane
measurements for Cl1358+62.

We determined the photometric parameters in two different ways, following
the procedures used by vDF. We first used Point-Spread Function images to
fit convolved $r^{1/4}$-law profiles to the galaxy images.  In addition, we
deconvolved the images with the {\sc clean} procedure (H\"ogbom 1974), and
derived growth curves for the galaxies. The results from both methods were
compared to estimate the errors. It is worth noting that the median
differences in $\re$ and $\mue$ were $+7.4\%$ and $-9.7\%$, but the
combined parameter $r_eI_e^{0.82}$ only differed by $-1.2\%$ for Cl1358+62
and $-1.4\%$ for MS2053--04. This is the combination of parameters that
enters the Fundamental Plane, and as a result, our subsequent analysis is
insensitive to the individual errors in $r_e$ and $I_e$.\footnote{ The same
applies for previous measurements at intermediate and low redshift
(J\o{}rgensen, Franx \& Kj\ae{}rgaard 1995a).} Because the Coma data were
derived from growth curves, we proceeded to use the growth curve results in
the following analysis. After calibration using Holtzman \etal\ (1995), the
photometry was transformed to the redshifted $V$-band, for direct
comparison to the Coma photometry. This is possible, because we have
observations in multiple passbands close to the redshifted $V$-band. Colors
were measured within an aperture of $r < 3r_e$, and Galactic extinctions
were derived from Burstein and Heiles (1982) and Cardelli, Clayton \&
Mathis (1989).

\subsection{Errors}

Errors have been determined directly from the spectroscopic, structural and
photometric fits, as well as the photometric transformations. The random
errors are listed in Table \ref{tab:params}, and the typical random error
bars are shown in Figure \ref{fpz} (as thin error bars). We have considered
possible sources of systematic errors and we have estimated their
contribution (shown as thick error bars in Figure \ref{fpz}). We have
several sources of systematic errors: (i) photometric transformations at
$\pm 0.05$ mag, which is dominated by the uncertainties in the absolute
zeropoint of the F814W passband; (ii) velocity dispersions, where our
procedures may have relative errors of $\pm 3\%$ with respect to similar
measurements at low redshift (the absolute velocity dispersions may be in
error systematically by up to $\pm 5$-7\%); (iii) structural parameters
($\re$ and $\mue$), where deviations from an \rq\ model could cause
systematic errors on the level of $\pm 1\%$ in the combined parameter $\re
\mue^{0.82}$.

Another source of uncertainty is due to departures from homology (see, {\it
e.g.\/}, Capelato \etal\ 1995, Ciotti \etal\ 1996). Non-homology can affect
our measurement of evolution through the aperture correction for the
velocity dispersions. Jorgensen \etal\ (1995b) determined the aperture
corrections empirically, by using long slit data on nearby galaxies. They
found no strong effect out to an effective radius. Therefore, these aperture
corrections are likely to be appropriate for most of our galaxies. However,
for the smallest galaxies, this correction is more uncertain and may
require future observations of velocity dispersion profiles to large radii
in a broad sample of nearby galaxies ({\it e.g.\/}, Corollo \etal\ 1995).


\section{The Fundamental Plane in Cl1358+62 and MS2053--04}

The Fundamental Plane for the clusters are shown in Figure \ref{fpz} along
with the FPs for Cl0024+16 (vDF) and Coma (JFK96). We use the coefficients
for the FP determined by JFK96 from a large sample of 225 early-type
galaxies in ten nearby clusters.  The figure shows clearly that a well
defined Fundamental Plane exists, {\it despite the fact that the galaxies
in the intermediate redshift clusters were chosen without morphological
information}. Furthermore, the sample is large enough to derive the scatter
about the Coma Fundamental Plane. We find surprisingly low {\it rms\/}
scatters in $\log \re$ of $\pm 0.064$, $\pm 0.065$, $\pm 0.060$, and $\pm
0.072$ for Coma, Cl1358+62, Cl0024+16, and MS2053--04, respectively. The
galaxies also show a large offset from the Coma relation, due mainly to
cosmological surface brightness dimming.

One interesting question is whether the coefficients of the FP are the same
in higher redshift clusters, {\it i.e.\/}, are the luminous and less
luminous galaxies evolving at the same rate? However, the current sample is
too small to provide a definitive answer. The weak indication that the
slope is flatter when the distant galaxies are taken together needs to be
verified with larger samples before any conclusions should be made (see
also vDF).

We determined the mean $M/L_V$ ratio for each cluster directly from the
Fundamental Plane zeropoint, adopting the slopes of the Fundamental Plane
of JFK96 and $q_0=0.05$. The resulting evolution of $M/L_V$ ratio is shown
in Figure \ref{mlz}. The errors are taken from \S 2.3 and have been added
in quadrature. Weighting the individual galaxies by their random errors
does not change the results significantly.

Clearly, the $M/L_V$ ratio is lower at higher redshift, consistent with
evolution of the stellar populations. We have drawn simple, single-burst
model predictions in the same plot, adopting formation redshifts $\zform$
of infinity, and $\zform=1$. The current data are not consistent with the
predictions for co-eval populations which have formed recently. More data
are needed to test whether more complex models with recent star formation
can be accommodated (see, e.g., Franx and van Dokkum 1996 and Poggianti \&
Barbaro 1996).


\section{Discussion}

We have measured structural parameters and central velocity dispersions for
galaxies in two clusters at intermediate redshift, Cl1358+62 at $z=0.33$
and MS2053--04 at $z=0.58$. The Fundamental Plane relations in the
intermediate redshift clusters are very similar to that found in Coma. This
observation demonstrates that mature early-type galaxies existed in these
clusters at $z\approx 0.6$; their primary epochs of star formation must
have occurred at much higher redshifts.

The sample is also large enough to measure the scatter in the Fundamental
Plane relation reliably. We find it to be low:  $\pm 0.067$ in $\log \re$,
or $\pm 17\%$ in $M/L_V$. This suggests that the populations are very
homogeneous, and that the age differences between the galaxies are not very
large (Ciotti, Lanzoni, \& Renzini 1996).

The mean $M/L_V$ ratio of the galaxies was clearly lower several Gyr ago,
consistent with passively evolving stellar populations. This evolution
depends on the formation redshift(s) of the population, the IMF(s), and
cosmological model ({\it e.g.\/}, Franx 1995). We show model predictions
for formation redshifts of $z=1$ and $z=\infty$ in Figure \ref{mlz}
(Tinsley \& Gunn 1976) using $q_0=0.05$. The new data are fully consistent
with a high formation redshift, $z_{form} > 2$, strengthening the
conclusion of vDF. This result is a lower limit; the constraints are even
stronger if $q_0=0.5$.

This interpretation, however, is complicated by the fact that mergers,
interactions, starbursts, and other processes may continue to transform
late-type galaxies into early-type galaxies: the early-type galaxies we
observe at high redshift may only be a subset of the early-type galaxies we
observe at low redshift. In this case, the early-type galaxies observed at
high redshift (if they remained undisturbed until the present) should be
compared to the oldest early-type galaxies locally. In some sense, these
high redshift early-type galaxies have been compared to a {\it mean\/} low
redshift counterpart, probably not as old as the comparison requires. Thus,
the formation redshift estimated from the $M/L$ evolution can be biased
upwards (see, {\it e.g.\/}, Franx and van Dokkum 1996).

We included two Cl1358+62 ``E+A'' galaxies in our sample to test whether
these galaxies are progenitors of early-type galaxies in low redshift
clusters.\footnote{These galaxies have spectra which can be interpreted as
a superposition of an old population (the ``E'' for early type), and a
young population with the spectrum of an A star (Dressler and Gunn 1983).}
They are shown in Figure \ref{fpz} by the ``x'' symbols. We can use the
Coma Fundamental Plane to measure the $M/L_V$ ratios of these ``E+A''
galaxies, assuming that their structural properties are similar to nearby
early-type systems. This assumption may not necessarily be correct, as the
Franx (1993) analysis of an E+A in Abell 665 ($z=0.18$) shows that it is
essentially bulgeless; it is not clear whether it will become an S0 or
remain a spiral system.

With this caveat in mind, we show the mean $M/L_V$ offset for the two E+A
galaxies in Figure \ref{mlz} as an ``x.'' This $M/L_V$ is consistent with a
``formation'' redshift of $z_f \approx 0.5$, a $V$-band luminosity weighted
mean of the formation redshifts of the subcomponents. This is consistent
with the hypothesis that the E+As have undergone a burst of star formation
1-2 Gyr before they have been observed (Dressler and Gunn 1983). We
conclude that a fraction of nearby cluster early-type galaxies has
undergone secondary bursts of star formation at $z<1$.

At this point, more work is needed to assess the relevance of the E+A
galaxies to the evolution of early-type galaxies and we will study the E+As
in these clusters in greater detail elsewhere. Observations of higher
redshift clusters will be needed to test whether star formation and
starbursts were even more prevalent at earlier times, as suggested by Lubin
(1996) and Rakos \& Schombert (1995). Our sample can also be compared to
surveys of nearby field galaxies, in which the data of Gonz\'alez (1993)
and Faber \etal\ (1995) suggest that a large age spread ($\sim 2$-18 Gyr)
exists. Thus, a large fraction of these experienced bursts of star
formation at $z<1$. Our data suggest that it might be hard, but not
impossible, to model the evolutionary history of the cluster galaxies in
the same way as the field galaxies.

\acknowledgements

We appreciate the effort of all those in the HST program that made this
unique Observatory work as well as it does. The assistance of those at
STScI who helped with the acquisition of the HST data is gratefully
acknowledged. We also appreciate the effort of those at the Keck, MMT and
KPNO telescopes who developed and supported the facility and the
instruments that made this program possible. Support from STScI grants
GO05989.01-94A, GO05991.01-94A, and AR05798.01-94A is gratefully acknowledged.

\clearpage
\section*{Figure Captions}

\figcaption{[Plate 1]
(a) Montage of color images for the sample galaxies in Cl1358+62, using
F606W and F814W; and in (b) MS2053--04 using F702W and F814W; (c) Spectrum
of galaxy \#256 in Cl1358+62; and (d) \#197 in MS2053--04.
\label{mosaic}}

\figcaption[show3.epsi]{
The FP for the four clusters with the mean FP for Coma (JFK96) plotted on
each panel. Note that the two E+As in Cl1358+62, shown as ``x''s, lie to
the left of the mean relation of the ``old'' Cl1358+62 early-type galaxies.
The previous results of JFK96 and vDF are shown as open circles. The new
data are shown as filled circles. Typical random (thin) and systematic
(thick) errors are shown.
\label{fpz}}

\figcaption[evolmod.epsi]{
The mean $M/L_V$ offsets with redshift, for $q_0=0.05$. The area enclosed
by solid lines corresponds to single burst models with $z_f=\infty$ and a
range of \imf{}s. The region marked by dash lines corresponds to the
equivalent $z_f=1$ models. The previous results of JFK96 and vDF are shown
as open circles. The new data are shown as filled circles. The ``x'' marks
the $M/L_V$ offset derived from the two E+A galaxies in Cl1358+62. The
errors were estimated by adding the random and systematic errors in
quadrature.
\label{mlz}}

\clearpage

\begin{deluxetable}{l c c l c c}
\footnotesize
\tablewidth{0pt}
\tablecaption{Fundamental Plane Parameters\label{tab:params}}
\tablehead{
\colhead{ID} &
\colhead{$I$} &
\colhead{$R-I$} &
\colhead{$\sigma$} &
\colhead{$r_e$} &
\colhead{$\mu_{e,V}$}\nl
\colhead{} &
\colhead{(mag)} &
\colhead{(mag)} &
\colhead{($\rm km\ s^{-1}$)} &
\colhead{(arcsec)} &
\colhead{($\rm mag\ arcsec^{-2}$)}\nl
\colhead{(1)}  &
\colhead{(2)}  &
\colhead{(3)}  &
\colhead{(4)}  &
\colhead{(5)}  &
\colhead{(6)}
}
\startdata
200&18.70&0.68&$135\pm 11$&0.913&22.50\nl
236&17.98&0.80&$166\pm 11$&0.541&21.93\nl
256&17.56&0.78&$273\pm \,\,\,7$&1.024&21.81\nl
269&17.85&0.82&$342\pm 10$&0.826&21.55\nl
298&18.25&0.82&$280\pm \,\,\,8$&0.448&20.77\nl
328&19.08&0.60&$\ \,98\pm \,\,\,7$&0.712&22.24\nl
375&17.44&0.81&$301\pm 11$&3.910&23.83\nl
408&19.04&0.77&$265\pm 17$&0.287&20.74\nl
454&18.71&0.73&$171\pm \,\,\,6$&0.780&22.23\nl
470&18.41&0.79&$185\pm \,\,\,6$&0.738&22.23\nl
311&20.52&1.12&$223\pm 25$&0.402&22.53\nl
197&18.59&1.20&$319\pm 18$&2.182&23.99\nl
422&20.59&1.20&$158\pm 22$&0.413&22.57\nl
551&20.89&1.14&$217\pm 19$&0.157&20.93\nl
432&20.41&1.08&$161\pm 20$&0.483&22.82\nl
\enddata
\tablecomments{
(1) Galaxy identification; 
(2) Integrated $I$ magnitude ($\pm0.1$ mag);
(3) $R-I$ color within a $3r_e$ aperture ($\pm 0.05$ mag);
(4) Velocity dispersion with formal (random) errors;
(5) Effective radius $r_e$ from {\sc clean}/growth curve fit of $r^{1/4}$-law to HST images;
(6) Surface brightness in rest frame $V$-band from that fit at $r_e$}
\end{deluxetable}

\clearpage

\centerline{\epsfxsize=8.8cm \epsfbox{figure2.epsi}}

\clearpage

\centerline{\epsfxsize=8.8cm \epsfbox{figure3.epsi}}


\clearpage

\begin{thebibliography}{}

 \bibitem{ab}Broadhurst T. J., Ellis R. S., \&  Shanks T. 1988, \mnras,
	235, 827
 \bibitem{bh}Burstein D., \&  Heiles C. 1982, \aj, 87, 1165
 \bibitem{ac}Butcher H., \&  Oemler A. 1978, \apj, 219, 18
 \bibitem{ad}Butcher H., \&  Oemler A. 1984, \apj, 285, 426
 \bibitem{cz}Capelato, H.~V., de Carvalho, R.~R., \& Carlberg, R.~G.
	1995, \apj, 451, 525
 \bibitem{cc}Cardelli, J.~A., Clayton, G.~C., \& Mathis, J.~S. 1989,
	\apj, 345, 245
 \bibitem{ae}Ciotti, L., Lanzoni, B., \& Renzini, A. 1996, \mnras, 282, 1
 \bibitem{fe}Corollo, C.~M., de Zeeuw, P.~T., van der Marel, R.~P.,
	Danziger, I.~J., \& Qian, E.~E. 1995, \apjl, 441, L25
 \bibitem{ae}Couch W. J., \& Sharples R. M. 1987, \mnras,
	229, 423
 \bibitem{af}Djorgovski S., \&  Davis M. 1987, \apj, 313, 59
 \bibitem{ag}Dressler A., \&  Gunn J. E. 1983, \apj, 270, 7
 \bibitem{ah}Dressler A., Lynden-Bell D., Burstein D., Davies R. L.,
	Faber S. M., Terlevich R. J., \& Wegner G. 1987, \apj, 313, 42
 \bibitem{ai}Ellis R. S., Smail, I., Dressler, A., Couch, W.J.,
	Oemler, A., Butcher, H., \& Sharples, R.M. 1996, preprint
 \bibitem{aj}Faber S. M., Dressler A., Davies R. L., Burstein D.,
	Lynden-Bell D., Terlevich R., \&  Wegner G. 1987,
        Faber S. M., ed., Nearly Normal Galaxies. Springer, New
	York, p. 175
 \bibitem{al}Faber, S. M., Trager, S. C., Gonz\'alez J. J., \& Worthey, G.
	1995, IAU Symp. 164, Stellar Populations (Dordrecht: Kluwer), 255
 \bibitem{an}Franx M. 1993, \pasp, 105, 1058
 \bibitem{an}Franx M. 1995 IAU Symp. 164, Stellar Populations
	(Dordrecht: Kluwer), 269
 \bibitem{ao}Franx M., \& van Dokkum, P.~G. 1996, {\it New Light on
	Galaxy Evolution\/}, eds. Bender, R., \& Davies, R.~L., in press.
 \bibitem{ap}Franx M., \etal\, 1997, in preparation
 \bibitem{aq}Gonz\'{a}lez, J.~J. 1993, Ph.D. Thesis, Univ. California,
	Santa Cruz
 \bibitem{ar}H{\" o}gbom J.~A. 1974, A\&AS, 15,417
 \bibitem{ho}Holtzman, J.~A., Burrows, C.~J., Casertano, S., Hester, J.~J.,
	Trauger, J.~T., Watson, A.~M., \& Worthey, G. 1995, \pasp, 107, 1065
 \bibitem{as}J\o{}rgensen I., Franx M., \& Kj\ae{}rgaard P. 1993, \apj,
	411, 34
 \bibitem{at}J\o{}rgensen I., Franx M., \& Kj\ae{}rgaard P. 1995a,
	\mnras, 273, 1097
 \bibitem{at}J\o{}rgensen I., Franx M., \& Kj\ae{}rgaard P. 1995b,
	\mnras, 276, 1341
 \bibitem{au}J\o{}rgensen I., Franx M., \& Kj\ae{}rgaard P. 1996,
	\mnras, 280, 167 [JFK96]
 \bibitem{aw}Kron, R. 1980, \apjs, 43, 305
 \bibitem{ax}Le F\'evre, O., \etal\, 1995, \apj, 461, 534
 \bibitem{ay}Lilly S.~J., Tresse L., Hammer F., Crampton D., \&
	Le F\`evre O. 1995, \apj, 455, 50
 \bibitem{bz}Lubin, L. 1996, \aj, accepted for publication
 \bibitem{pz}Poggianti, B. M., \& Barbaro, G. 1996, \aap, in press
 \bibitem{cz}Rakos, K.~D., \& Schombert, J.~M. 1995, ApJ, 439, 47
 \bibitem{tg}Tinsley, B.~M., \& Gunn, J.~E. 1976, \apj, 203, 52
 \bibitem{bb}van Dokkum, P.~G., \& Franx M. 1996, \mnras, 281, 985 [vDF]
 \bibitem{bc}Vogt, N.~P., Forbes, D.~A., Phillips, A.~C.,
	Gronwall, C., Faber, S.~M., Illingworth, G.~D., \& Koo, D.~C.
	1996, \apjl, 465, L15


\end{thebibliography}
\end{document}